\documentstyle[psfig]{aipproc}

\def\bg#1{\mbox{\boldmath$#1$}}

\newcommand{\del}{\partial}
\newcommand{\beq}{\begin{eqnarray}}
\newcommand{\eeq}{\end{eqnarray}}
\newcommand{\be}{\begin{eqnarray*}}
\newcommand{\ee}{\end{eqnarray*}}
\newcommand{\bk}{{\bf k}}
\newcommand{\bp}{{\bf p}}
\newcommand{\bq}{{\bf q}}

\newcommand{\bx}{{\bf x}}

\newcommand{\ra}{\rightarrow}
\newcommand{\e}{\epsilon}

\newcommand{\nn}{\nonumber}

\def\square{\vcenter{\vbox{\hrule height.4pt
          \hbox{\vrule width.4pt height6pt
          \kern6pt\vrule width.4pt}\hrule height.4pt}}}
\def\boxx{\square}

\begin{document}

\title{Applications of Effective Lagrangians}
\author{Finn Ravndal$^*$\footnote{Invited talk at {\it Beyond the
       Standard Model V}, Balholm, Norway, April 29 - May 4, 1997}
\address{$^*$Institute of Physics, University of Oslo, N-0316 Oslo, Norway}}
\maketitle

\begin{abstract}
Effective Lagrangians were originally used only at the tree level as 
so-called phenomenological Lagrangians since they were in general 
non-renormalizable. Today they are treated as effective field 
theories valid below a characteristic energy scale. Quantum corrections can
then be calculated in a consistent way as for any renormalizable theory.
A few applications of the Euler-Heisenberg Lagrangian for interacting photons
at low energies are presented together with recent developments in the use of 
QED for non-relativistic systems. Finally, the ingredients of 
an effective theory for the electroweak sector of the Standard Model are discussed
in the case of a non-linear realization of the Higgs mechanism using 
the St\"uckelberg formalism.
\end{abstract}

\section*{Introduction}
As a graduate student at Caltech in the early seventies when the Standard Model was 
established, we were surprised to notice that Richard Feynman did not take very much
interest in unified theories of weak and electromagnetic interactions. When he was
asked why, he said that these theories were all built on the principle of being
renormalizable. And since nobody understood what renormalization {\it really} was,
one were not allowed to construct new physical theories based on such a principle.

However, at the very same time Ken Wilson was unraveling the physics behind renormalization,
although in the somewhat separate field of critical phenomena. His deep insight led to the
understanding of any field theory being directly tied to the scale of the phenomena it
is supposed to describe. If we have a theory at a certain scale, we can from it
construct a new, {\it effective} field theory at a larger scale involving new fields and
interactions which in principle can be calculated from the theory at the shorter
scale. Going in the opposite direction from a large to a smaller scale, there is
no reason why the same theory should be applicable. So field theories in general should
be non-renormalizable. In the special case where one can go to a much smaller scale with the 
same theory, the theory is renormalizable. But that is the exceptional case.
  
In such a short talk it is impossible to say much about the
very many applications of effective Lagrangians in modern high energy physics.
More detailed reviews have been given by Kaplan\cite{finnr:Kaplan} and 
Manohar\cite{finnr:Manohar}. I have here chosen first to describe the illustrative 
example involving the interactions of low energy photons based upon the 
Euler-Heisenberg effective Lagrangian.
Including non-relativistic electrons one can then use NRQED to calculate
radiative and relativistic corrections in atomic physics to very high accuracy. 
Finally, I'll say a few words about the possibility of standard, electroweak 
interactions also being described by an effective theory without an
elementary Higgs boson valid below a characteristic energy scale where new, unknown physics appears.

I will not have time to discuss the surprising and beautiful results coming out
of the marriage of quantum theory and gravitation. Usually this is said not to work and to
represent one of the remaining big, unsolved problems in modern physics. Again the reason has
been that Einstein's geometric Lagrangian is a non-renormalizable theory. But now 
John Donoghue\cite{finnr:Donoghue} has shown that by treating it as an effective theory, 
quantum corrections can be calculated in a consistent way for all energies below the 
Planck scale. The lowest order quantum correction to the gravitational attraction between 
two masses has been calculated and corresponds
to Newton's constant effectively becoming smaller at shorter distances. This is just as to
be expected in a non-abelian gauge theory. In almost every thinkable astrophysical situation 
the quantum effects are completely negligible. At ordinary energies and curvatures this will 
not change even if we should get a more fundamental theory of gravity valid at shorter scales.

\section*{Euler-Heisenberg interaction}
QED describes the interactions between the electromagnetic field $A_\mu(x)$ and the
electron field $\psi(x)$ by the standard Lagrangian
\be
   {\cal L}(A,\psi) = -{1\over 4}F_{\mu\nu}^2 
                    + \bar{\psi}(\gamma^\mu(i\del_\mu - eA_\mu) - m)\psi 
\ee
where $F_{\mu\nu} = \del_\mu A_\nu - \del_\nu A_\mu$ is the field strength.
With no matter present and for energies much below the electron mass $m$, the
photons can interact only indirectly via virtual electron loops in the
vacuum. These interactions are suppressed by inverse powers of the electron 
mass compared with the Maxwell term and are thus very small. The effective
Lagrangian should be Lorentz and gauge invariant and respect parity
invariance.
\begin{figure}[htb]
     \begin{center}
     \mbox{\psfig{figure=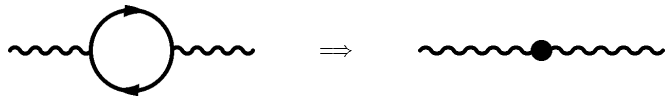,height=12mm}}
     \end{center}
     \caption{\protect The low-energy Uehling interaction from the vacuum
                       polarization loop.}       \label{fig1}
\end{figure}
It can thus be constructed only from the field strength $F_{\mu\nu}$ and
its derivatives. In this way we can write down the form of the effective
Lagrangian involving the lowest, dimension six operator,
\beq
     {\cal L}_U = -{1\over 4}F_{\mu\nu}^2 
                      + C_U{1\over m^2}F_{\mu\nu}\boxx F^{\mu\nu}  \label{fr:LU}
\eeq
where $\boxx \equiv \del_\mu\del^\mu$. This new term is just the Uehling interaction
modifying the photon propagator at low energies\cite{finnr:Uehling}. To lowest order in the fine
structure constant $\alpha = e^2/4\pi$ the new coupling constant is
$C_U = \alpha/60\pi$ resulting from the Feynman diagram in Figure 1.

However, in the absence of matter this new term will not contribute to any physics.
This is most easily seen by using the equation of motion for the free field
which is simply $\boxx\, F_{\mu\nu} = 0$ and makes the Uehling term go away. 
Higher order operators in the effective Lagrangian will then contribute instead.
Their coefficients must be found by matching to the underlying theory which
is here QED. This is equivalent to integrating out the electron field in the
Lagrangian\cite{finnr:IZ}. The first non-trivial photon interaction
is obtained with dimension eight,
\beq
     {\cal L}_{EH} = -{1\over 4}F_{\mu\nu}^2  
                    +  {\alpha^2\over 90m^4}\left[(F_{\mu\nu}F^{\mu\nu})^2 
                    + {7\over 4}(F_{\mu\nu}\tilde{F}^{\mu\nu})^2\right]    \label{fr:LEH}    
\eeq
where $\tilde{F}_{\mu\nu} = {1\over 2}\epsilon_{\mu\nu\rho\sigma}F^{\rho\sigma}$.
This is the Euler-Heisenberg Lagrangian\cite{finnr:EH} giving a non-linear interaction
between photons. 
\begin{figure}[htb]
     \begin{center}
     \mbox{\psfig{figure=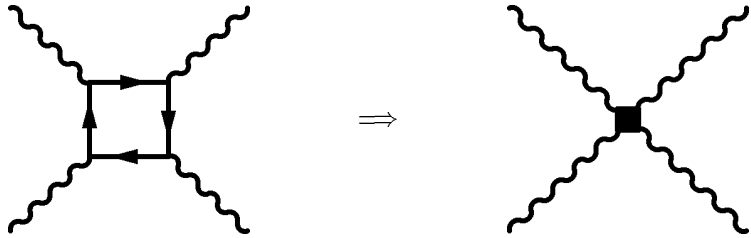,height=30mm}}
     \end{center}
     \caption{\protect The Euler-Heisenberg interaction between four photons.}       \label{fig2}
\end{figure}
At the microscopic scale it is caused by the coupling of 
the four photons to a virtual electron loop as shown in Figure 2. Higher order terms 
in the expansion will represent interactions between more photons.

We will here mention just two applications of this effective Lagrangian. It was already used
by Euler in 1936 to derive the elastic photon-photon cross section at the tree level. 
Even if it represents a non-renormalizable theory, Halter\cite{finnr:Halter} 
showed that a finite result
for the one-loop correction to the scattering amplitude can be calculated from 
the diagram in Figure 3 and its crossed versions. Just from dimensional arguments we know that
the correction must be order of magnitude $\alpha^2 (\omega/m)^4$ smaller 
than the tree-level amplitude where $\omega$ is the photon energy. 
\begin{figure}[htb]
     \begin{center}
     \mbox{\psfig{figure=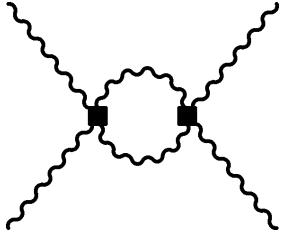,height=30mm}}
     \end{center}
     \caption{\protect Quantum correction to photon-photon scattering.}       \label{fig3}
\end{figure}
However, this is not the lowest radiative correction to the Euler cross section. If the
effective Lagrangian (\ref{fr:LEH}) is derived from QED more accurately, i.e. to two-loop
order, the four-photon couplings will be modified by terms of order $\alpha$. These have
been calculated by Ritus\cite{finnr:Ritus},
\beq
     (FF)^2 &\ra & \left(1 + {40\over 9}{\alpha\over\pi}\right) (FF)^2\\            \nn
     (F\tilde{F})^2 &\ra & \left(1 + {1315\over 252}{\alpha\over\pi}\right)(F\tilde{F})^2
\eeq
and recently confirmed by Reuter, Schmidt and Schubert\cite{finnr:RSS}. Obviously, the resulting 
corrections to the cross section are much larger than the contribution obtained by Halter.
This shows that if one wants to obtain correct results of a fundamental theory using a
low-energy effective theory, they have to be properly {\it matched} to the appropriate order 
in the fundamental coupling constant. 

Free photons at finite temperature $T$ exerts a pressure given by the Stefan-Boltzmann formula 
$P = (\pi^2/45)T^4$.
In full QED they will interact with the virtual electrons in the vacuum. For temperatures
$T < m$ the dominant interaction is given by the Euler-Heisenberg
Lagrangian (\ref{fr:LEH}).
\begin{figure}[htb]
     \begin{center}
     \mbox{\psfig{figure=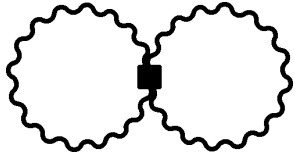,height=20mm}}
     \end{center}
     \caption{\protect Correction to the Stefan-Boltzmann free energy in the
                        effective theory.}       \label{fig4}
\end{figure}
Again we see from dimensional reasoning that the correction to the pressure will be of the order
$\alpha^2(T/m)^4$. It can be obtained from the Feynman diagram in Figure 4 which gives for the 
total photon pressure\cite{finnr:KR}
\beq
     P = {\pi^2\over 45}T^4 + {22\pi^4\alpha^2 \over 3^5\, 5^3} {T^8\over m^4}  
\eeq
In full QED it will follow from the evaluation of a difficult three-loop diagram, while using 
effective field theory it is obtained essentially from a one-loop diagram.
The above result has previously been obtained by Barton\cite{finnr:Barton} using 
semi-classical methods. Since the Uehling term is absent in the effective 
Lagrangian, we see that there is no term $\propto T^6$ in the result. This is 
also the case for the pressure of massless pions in chiral theory\cite{finnr:Gerber}.

\section*{Non-relativistic QED}
When there are also non-relativistic electrons interacting with
the photons, covariant and renormalizable QED can be replaced by a
non-relativistic and non-renormalizable effective theory called NRQED 
as first proposed by Caswell and Lepage\cite{finnr:CL}. Relativistic effects
are represented by local operators of higher dimensions. Since the relevant, 
low-energy degrees of freedom are separated out in the initial, effective Lagrangian, 
it allows a cleaner and more systematic derivation of higher order radiative
corrections. Divergent loop integrations are cut off at a momentum $\Lambda \le m$. 
The higher order coupling constants will in general also involve this cutoff
and are most conveniently obtained by matching QED and NRQED scattering amplitudes.
With the values of the effective couplings established this way, they can then
also be used for bound states. The over-all $\Lambda$-dependence will then drop
out of the final, physical result. In this way it can be used to calculate radiative
shifts of energy levels in simple atoms with high precision\cite{finnr:KN}. There is no longer
any need for the covariant Bethe-Salpeter formalism for bound states. Instead
one can use the standard Schr\"odinger wave functions found in any textbook
on quantum mechanics.

The general structure of the NRQED Lagrangian can be derived from the requirement
of being invariant under gauge and Galilean transformations. Many of the terms
can be obtained directly from the Dirac equation in the non-relativistic limit.
The leading part is the Pauli Lagrangian
\beq
    {\cal L}_{Pauli} = \psi^\dagger\left[iD_0 + {1\over 2m}{\bf D}^2 
                     + {e\over 2m}{\bg\sigma}\cdot{\bf B}\right]\psi       \label{fr:Pauli}
\eeq
for a non-relativistic electron interacting with the photon field $A^\mu = (\Phi, {\bf A})$.
Here we have introduced the covariant derivatives $D_0 = \del_0 + ie\Phi$ and ${\bf D} = 
{\bg\nabla} - ie{\bf A}$ in the Coulomb gauge where ${\bg\nabla}\cdot{\bf A} = 0$ and 
${\bf B} = {\bg\nabla}\times {\bf A}$ is the magnetic field. A photon with the four-momentum
$k^\mu = (\omega,\bk)$ has a static Coulomb propagator
propagator $D_{00}(\omega,\bk) = {1/\bk^2}$ while the transverse propagator is
\be
     D_{ij}(\omega,\bk) = {1\over k^2 + i\e}\left(\delta_{ij} - {k_ik_j\over\bk^2}\right)
\ee
Similarly, the propagator for a non-relativistic electron with energy $E$ and momentum 
$\bp$ is $ G(E,\bp) = 1/( E - {\bp^2/ 2m} + i\e)$.  There are three
couplings of the photon in this theory, an electric coupling to the electron charge
and a magnetic coupling to the electron spin plus a seagull coupling involving two photons.

\begin{figure}[htb]
     \begin{center}
     \mbox{\psfig{figure=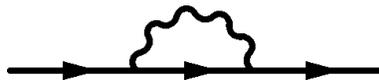,height=10mm}}
     \end{center}
     \caption{\protect Loop correction to the electron propagator due
                  to a transverse photon.}       \label{fig5}
\end{figure}
The lowest order propagator correction comes from the diagram in Figure 5. When both the
vertices are electric, it gives
\be
     \Sigma = \left({ie\over 2m}\right)^2 \int\!{d^4k\over (2\pi)^4}
              (2p_i - k_i)(2p_j - k_j)G(E - k_0,\bp - \bk)D_{ij}(k_0,\bk)
\ee
Integrating first over $k_0$ gives $k_0 = |\bk| = \omega$. The remaining integral
is linearly divergent.  After an angular integration, it then gives the self-energy
    $\Sigma = - (2\alpha/3\pi m^2)p^2 \Lambda$.           
It can be interpreted as a radiative correction to the electron mass. The same
diagram with one electric and one magnetic coupling gives zero, while two magnetic couplings
give similarly a cubic divergent result.

Next we consider the simplest vertex correction shown in Figure 6 for an electron scattering
off a potential $\Phi(\bx)$. If $\bp$ and $\bq$ are the initial and final electron momenta,
the momentum transfer from the potential is $\bk = \bq - \bp$. 
\begin{figure}[htb]
     \begin{center}
     \mbox{\psfig{figure=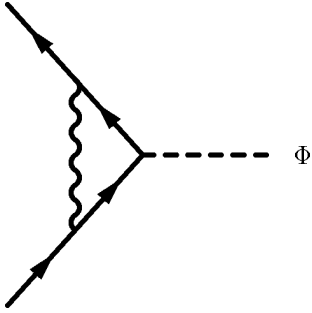}}
     \end{center}
     \caption{\protect Vertex correction in NRQED for scattering off a potential.}    \label{fig6}
\end{figure}
The linearly 
divergent corrections on the external electron legs are cancelled by the above self-energy.
There is a remaining, logarithmicly divergent wave function renormalization which combines
with a similar contribution from the non-relativistic vertex correction. They  both exhibit
an IR divergence which is cured by giving the photon a small mass $\lambda$ so 
that $\omega^2 = \bk^2 + \lambda^2$. We are then left with the net contribution 
\be
     V_{\bq\bp} = -{\alpha\over 3\pi m^2}(\bp - \bq)^2
                        \left(\log{2\Lambda\over\lambda} - {5\over 6}\right)e\Phi(\bq - \bp)
\ee
from these three diagrams.
The result is thus proportional with $\bk^2\Phi(\bk)$. In coordinate space it corresponds
to a higher order interaction ${\bg\nabla}\cdot{\bf E}$ where ${\bf E} = - {\bg\nabla}\Phi$
is the electric field. But this is just the Darwin interaction. It is needed as a counterterm
when the Pauli Lagrangian is used at the 1-loop level.

When one considers other scattering amplitudes, additional counterterms are needed. So
to this accuracy the non-relativistic Lagrangian (\ref{fr:Pauli}) has to be extended by including
these new interactions and it will take the form
\beq
   {\cal L}_{NRQED} &=& \psi^\dagger\left[iD_0 + {1\over 2m}{\bf D}^2 + C_K{1\over 8m^3}{\bf D}^4
                         + C_F{e\over 2m}{\bg\sigma}\cdot{\bf B}\right. \\ \nn
     &+&\left. C_D{e\over 8m^2}({\bf D}\cdot{\bf E} - {\bf E}\cdot{\bf D}) 
                    + C_{LS}{ie\over 8m^2}{\bg\sigma}\cdot
       ({\bf D}\times{\bf E} - {\bf E}\times{\bf D}) \right]\psi   \label{fr:nrqed}
\eeq
The most general effective Lagrangian to order $1/m^3$ has recently been presented by 
Manohar\cite{finnr:AM}.
Just as the Darwin term is a relativistic effect, the strength of these new interaction can 
be obtained from the non-relativistic reduction of the Dirac equation. The coupling constants
$C_i$ in the above effective Lagrangian are therefore all equal to one at tree level. 
To first order in $\alpha$ NRQED must now give the same results as obtained from 
renormalized QED. Again considering scattering off external potentials, this form of 
matching gives for the correction to the kinetic energy $C_K = 1$ while\cite{finnr:KN}
\be
    C_F &=& 1 + {\alpha\over 2\pi}  \\
    C_D &=& 1 + {8\alpha\over 3\pi}\left[\log{m\over 2\Lambda} - {3\over 8} + {5\over 6}\right]
              + {\alpha\over \pi} \\
    C_{LS} &=& 1 + {\alpha\over \pi}
\ee
The magnetic Fermi coupling has been renormalized so that the effective magnetic moment of the
electron at low energies now includes the anomalous contribution. Notice that the photon mass
$\lambda$ does not appear in the effective coupling constants. This is because the IR 
behaviour of both NRQED and QED should agree and thus $\lambda$ cancelles out in the 
matching\cite{finnr:KL}.
These renormalized couplings together with the Uehling modification (\ref{fr:LU}) of the photon 
propagator are now sufficient to give a much more direct derivation of the Lamb shift in
hydrogen than is usually found in textbooks\cite{finnr:lamb}.

\begin{figure}[htb]
     \begin{center}
     \mbox{\psfig{figure=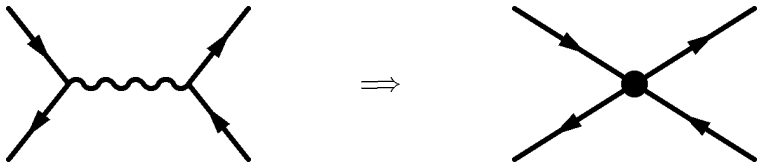}}
     \end{center}
     \caption{\protect Four-fermion vertex describing annihilation in
                              the effective theory.}    \label{fig7}
\end{figure}
Kinoshita and Nio have used NRQED to calculate the hyperfine splitting
in muonium to the highest precision\cite{finnr:KN}. 
Similarly, Labelle\cite{finnr:thesis} and collaborators\cite{finnr:LLM} have 
made applications to radiative shifts of energy levels and lifetimes of positronium. 
In lowest order one encounters here the four-fermion operator
$(\psi^\dagger{\bg\sigma}\chi)\cdot(\chi^\dagger{\bg\sigma}\psi)$ describing the annihilation
and subsequent creation of an electron $\psi$ and a positron $\chi$ via a photon
as shown in Figure 7. Evaluating the diagram, one finds the corresponding coupling
constant to be $\alpha\pi/m^2$. 
\begin{figure}[htb]
     \begin{center}
     \mbox{\psfig{figure=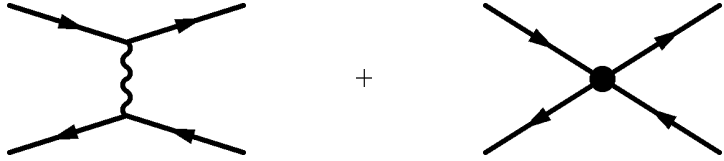,height=20mm}}
     \end{center}
     \caption{\protect Lowest order contributions to the hyperfine
                     splitting in positronium.}    \label{fig8}
\end{figure}
Since the virtual photon has the 
relativistic energy $2m$, the process is thus represented by a local operator in NRQED. 
Together with the ordinary Fermi interaction it will
contribute to the lowest hyperfine splitting $\Delta E$ between ortho- and para-positronium
as shown in  Figure 8. They give the textbook result $\Delta E = \alpha^4m(1/3 + 1/4)$.

When considering higher order interactions in positronium, additional
electron-posi\-tron contact terms will be needed. They will have the general spinor structure
\be
     {\cal L}_{ct} = C_1(\psi^\dagger\chi)(\chi^\dagger\psi) 
     + C_2(\psi^\dagger{\bg\sigma}\chi)\cdot(\chi^\dagger{\bg\sigma}\psi)
     + C_3(\psi^\dagger{\bf D}\chi)\cdot(\chi^\dagger{\bf D}\psi) + \ldots
\ee
when allowing for Fierz transformations. The effective coupling constants will in
general be complex and thus contribute both to the radiative shifts of energy levels and 
lifetimes. Again by matching to QED the full values of $C_1$ and $C_2$ to lowest order 
in $\alpha$ has recently been obtained\cite{finnr:LZB}. They reproduce the Karplus-Klein 
result for the hyperfine splitting in positronium to order $\alpha^5$\cite{finnr:KK}. 
An even more precise matching gives now the splitting to order $\alpha^6$ \cite{finnr:HLZ}.

\section*{Electroweak vector bosons}
In the Standard Model the electroweak bosons get masses from the spontaneous
breaking of the underlying $SU(2)_L\otimes U(1)_Y$ symmetry. When this is
realized in a linear way, we will have a renormalizable theory with an
elementary Higgs bosons. In principle, we then have a quantum field theory
describing essentially all fundamental phenomena ranging over scales varying 
roughly fifteen orders of magnitude. Something like this we have never seen in
physics before and the Standard Model would really have to be a very fundamental
theory. But with so many unknown parameters and with a rather ad hoc internal structure,
it is more likely that the model is just an effective theory valid up to some
unknown energy threshold where new physics appears. 
The requirement of being renormalizable is then not so pressing anymore, and 
the electroweak symmetry can be realized in a non-linear way with no 
elementary Higgs boson. But the would-be Goldstone bosons are still needed
in order to give masses to the electroweak vector bosons and allow a perturbative
evaluation of their quantum effects.

This was understood already in 1956 by St\"uckelberg\cite{finnr:Stuck}. Consider
the Lagrangian
\beq
     {\cal L}_V = -{1\over 4}F_{\mu\nu}^2 - {1\over 2}m^2V_\mu^2
                + a(V_\mu V^\mu)^2 + b(F_{\mu\nu}F^{\mu\nu})^2 + \ldots  \label{fr:LV}
\eeq
for an interacting, massive vector field $V_\mu$ where $F_{\mu\nu} = 
\del_\mu V_\nu - \del_\nu V_\mu$. Except for Lorentz and parity 
invariance, there are no other symmetries. It defines in general a non-renormalizable 
field theory which we now can treat as an effective theory. But
looking at the free propagator
\be
   \Delta_{\mu\nu}(k) = \left(\eta_{\mu\nu} - {k_\mu k_\nu\over m^2}\right)
                        {1\over k^2 - m^2 + i\e}
\ee
we see that is goes to a constant at very high momenta $k \gg m$ which makes
the evaluation of higher order loop diagrams impossible.

This technical problem can be cured according to St\"uckelberg by replacing
the massive vector boson $V_\mu$ by a massless vector boson $A_\mu$ plus
a massless scalar field $\theta$. From the covariant derivative
$D_\mu = \del_\mu - igA_\mu$, we can then replace the massive field with the 
composite field
\be
     V_\mu = {i\over g}U^\dagger D_\mu U = A_\mu - {1\over m}\del_\mu\theta
\ee
where $U(x) = \exp{(i\theta(x)/v)}$. The mass term in ${\cal L}_V$ then becomes
the kinetic term for the $\theta$ - field with $m = gv$. Now we have
$F_{\mu\nu} = \del_\mu A_\nu - \del_\nu A_\mu$ and the resulting theory of
these two massless fields has developed a new, local $U(1)$ gauge symmetry 
under which $\theta \ra \theta + v\chi$ and $A_\mu \ra A_\mu 
+ (1/g)\del_\mu\chi$. Via the  St\"uckelberg construction we thus see that the above
massive Lagrangian (\ref{fr:LV}) is just the strongly coupled Abelian Higgs model
in the unitary gauge.

This point of view was first introduced in the context of the Standard Model by
Cornwall, Levin and Tiktopoulos\cite{finnr:CLT} in 1974. Since then it was
advocated for  by Chanowitz, Georgi and Golden\cite{finnr:CGG} and others\cite{finnr:others}.
It has been used extensively by Yuan and collaborators\cite{finnr:Yuan} to consistently 
calculate radiative corrections
in the Standard Model without any elementary Higgs boson. Any coupling between
the physical vector bosons $A_\mu$, $Z_\mu$ and $W_\mu^\pm$ are then allowed
as long as it is invariant under Lorentz and local electromagnetic gauge
transformations. In this effective theory of electroweak interactions
anomalous couplings will therefore appear reflecting the existence of new 
physics at higher energies. They will offer a much richer world of 
physical phenomena than with a renormalizable Standard Model and a
fundamental Higgs particle.

\end{document}